# Introducing Development Features for Virtualized Network Services

Steven Van Rossem, Wouter Tavernier, Didier Colle and Mario Pickavet and Piet Demeester
Authors are with Ghent University - imec, IDLab, Department of Information Technology
iGent Tower - Department of Information Technology
Technologiepark-Zwijnaarde 15, B-9052 Ghent, Belgium
Email: {steven.vanrossem, wouter.tavernier, didier.colle, mario.pickavet, piet.demeester} @ugent.be

**Abstract** – Network virtualization and softwarizing network functions, are trends aiming at higher network efficiency, cost reduction and agility. They are driven by the evolutions in Software Defined Networking (SDN) and Network Function Virtualization (NFV). This shows that software will play an increasingly important role within telecommunication services, which were previously dominated by hardware appliances. Service providers can benefit from this, as it enables a faster introduction of new telecom services, combined with an agile set of possibilities to optimize and fine-tune their operations. However, the provided telecom services can only evolve if the adequate software tools are available. In this article, we explain how the development, deployment and maintenance of such an SDN/NFV-based telecom service puts specific requirements on the platform providing it. A Software Development Kit (SDK) is introduced, allowing service providers to adequately design, test and evaluate services before they are deployed in production and also update them during their lifetime. This continuous cycle between development and operations, a concept known as DevOps, is a well-known strategy in software development. To extend its context further to SDN/NFV-based services, the functionalities provided by traditional cloud platforms are not yet sufficient. By giving an overview of the currently available tools and their limitations, the gaps in DevOps for SDN/NFV services are highlighted. The benefit of such an SDK is illustrated by a secure content delivery network service (enhanced with deep packet inspection and elastic routing capabilities). With this use-case, the dynamics between developing and deploying a service are further illustrated.

# 1 Introduction

Modern-day telecom services show an increasingly dynamic behavior, causing network operators and service providers to adopt a more unified and elastic deployment approach. They move away from (vendor-) specific hardware middleboxes at centralized locations and instead use resource virtualization, distributed cloud-based platforms and global partnerships to respond efficiently to market demands. Economic viability requires high automation and scalability of resources, while still meeting stringent customer requirements such as: fast deployment, zero perceivable interruption and high personalization of services [1]. In this context, we investigate how to provide telco-grade solutions for the service development process.

## 1.1 The Evolution of SDN/NFV Development

A full-fledged development environment for NFV/SDN-based services builds upon the evolutions in three overlapping areas: programming languages or software tools in general, SDN/NFV related techniques and service platforms [2]. Figure 1 describes this evolution. NFV/SDN-based network services rely on a wide set of standards and technologies ranging from virtualization and network programming techniques to automation and monitoring tools. Auxiliary features help to deploy, configure and scale the service components in different infrastructure environments. An all-inclusive role is played by the Management and Orchestration (MANO) platforms, providing functions for automated deployment and operation of network services. This calls for dedicated support to adequately test and debug those service control mechanisms, before they are actually deployed in production.

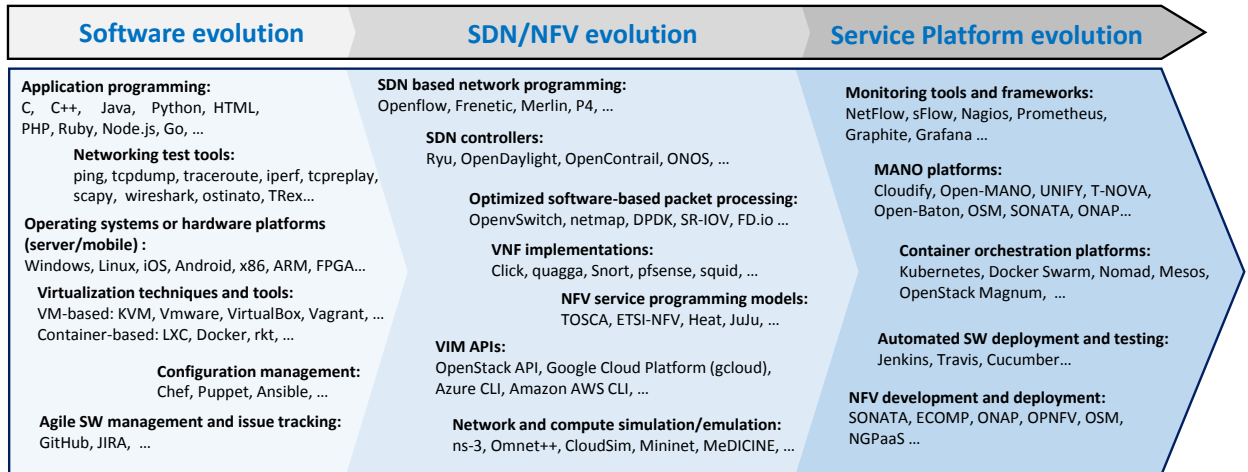

Figure 1: Evolution of the SDN/NFV eco-system. Growing functionality integration and abstraction require a growing set of dedicated tools for development, testing and debugging.

The next step would be to consolidate all these discrete NFV/SDN related tools into a unified SDK environment. But before we select the most interesting tools, let us have a deeper look into the specific components and characteristics of a modern network service.

## 1.2 Service Function Chaining

NFV and SDN are two main advancements that fundamentally change the way how network services are deployed. NFV aims at softwarizing (hardware-based) network functionalities like: packet filtering and forwarding, Network-Address-Translation (NAT), Quality-of-Service (QoS) management, WAN optimization… The new Virtualized Network Function (VNF) is now an isolated software 'image', ready to be deployed on generic, common-of-the-shelf servers. The infrastructure is now virtualized, enabling more fine-grained ways to consume compute, storage, and network resources. Complementary to NFV, SDN allows flexible and easier control of the networking between VNFs. The intelligence or algorithm which decides where traffic should be steered to, is implemented in a separate, over-looking control entity or SDN-controller. This control plane instructs the underlying packet forwarding devices or data plane using a well-defined protocol such as Openflow or NETCONF. The result is a centralized and programmable network management.

The flexibility provided by above described technologies, leads to the concept of Service Function Chaining (SFC). VNFs are deployed on infrastructure nodes which can be located at both the network edge and core. The telco-grade network service is now deployed as a chain of VNFs, dynamically connected into various topologies, as SDN provides ways to programmatically setup network links between several servers in one or more data centers. In the next section, we sketch a practical example of this.

### 1.2.1 Example Use-Case: Secure Content Delivery

To illustrate better the dynamics enabled by NFV, SDN and SFC, a typical telco service example is given in Fig. 2. It is a secured Content Delivery Network (CDN) where service subscribers can both download and upload content using their local user applications. Downwards, media can be streamed from a cloud VM which serves as a central database, or from Cache-VNFs which cache often streamed data in edge servers for faster delivery. Upwards, end-users push content to the cloud VM, by first passing through a Deep Packet Inspection (DPI)-VNF for security reasons. Both DPI and Cache VNFs are deployed in a distributed way, near the network edge, which enables better quality of experience by locating these functions closer to the (mobile) end-users. A Router-VNF aggregates the traffic before it reaches the cloud VM. The Router allows easier measurement of parameters like data volume, needed for billing or Service-Level-Agreement (SLA) monitoring. It isolates the service-dedicated routing process into a dedicated VNF. In case of flash crowds or other sudden data volume changes, the different VNFs can elastically scale, in function of the needed packet rate. We will use this use-case throughout the article to illustrate how SDN/NFV-based services are deployed and how this impacts their development environment.

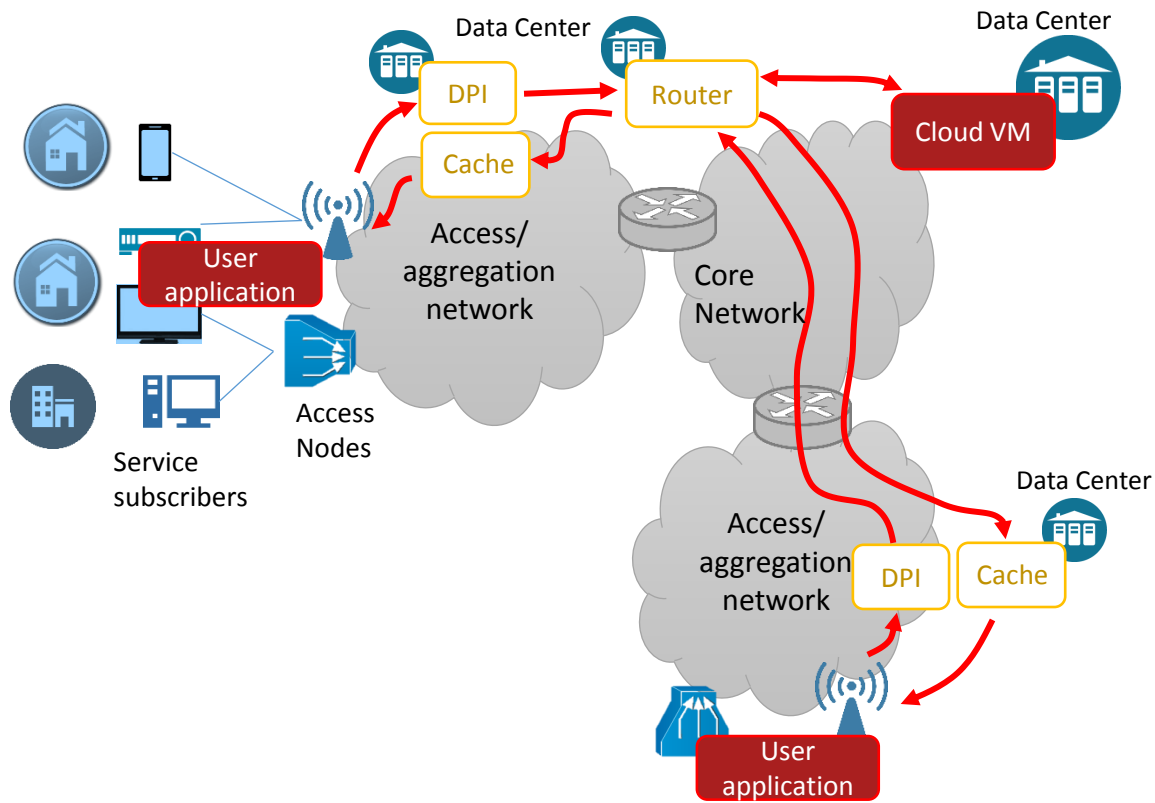

Figure 2. Example of an SDN/NFV-based Service Function Chains providing a secure CDN where each network function is implemented as a VNF.

## 1.3 Cloud Application vs. SDN/NFV-based Service

We want to highlight that SDN/NFV-based telecom services extend the classic cloud applications to many additional domains. Cloud applications are generally application-layer based, with a three-tier architecture consisting of a web-, application- and database server in the backend. A local or mobile device is used in the frontend. Moreover, cloud application software is typically not tuned for rapid lower-layer packet processing, needed in for example DPI, Cache or Router VNFs. Instead, cloud apps focus on endpoint functionality like web applications. The generic cloud eco-system is basically a connection from user to data center, where the network infrastructure in between is not leveraged. This is different from the SFC example in Fig. 2, where orchestration also needs to reach the access and core networks. A wider orchestration domain enables better optimization regarding the placement and resource use of the network functions, especially in case of large-scale services where many users are distributed over multiple access networks. It is envisioned that next generation of telecom services will rely heavily on dynamic service chains in the provider networks [1].

Cloud applications have limited, often single data center, orchestration possibilities. Therefore only sub-optimal scaling strategies can be used in certain cases. Typically this means cloning VNF images and putting a load-balancer in front of them, or adding more resources to the VNF like CPU and memory. In high-speed NFV-based services, adding a load-balancer is not always optimal, because a simple load-balancing action might have a processing cost in the same order as the original packet-handling itself, thus not leading to any improved processing speed. Instead, placing the VNFs closer to the edge might prove a better solution, as shown with the DPI and Cache VNFs in Fig. 2. During scaling, stateful VNFs might also require a more intelligent state migration strategy instead of simple cloning. The foregoing indicates that pre-defined auto-scaling and data-analytics provided by the operator do not always unlock the full potential of NFV/SDN-based services. Customized and service-specific actions defined by the service developer can handle certain lifecycle events more optimally by controlling scaling and placement mechanisms more closely [3, 4].

In the remainder of this article, section 2 discusses the different actors and position of the SDK in the telco-grade eco-system. In section 3, we detail the necessary features for the SDK environment to support all aspects of the SDN/NFV-based network service in practice. We conclude in section 4 with an overview of the SDK's challenges and opportunities.

## 2 The Service Development and Deployment Process

To understand the requirements for the SDK environment, we give an overview of the deployment process of NFV/SDN-based services, as presented in Fig. 3. We categorize three main groups of stakeholders in the service's lifecycle: **Vendors** or **Service Developers** use the SDK to create or edit services and package them, ready to deploy. The **Operator** or **Service Provider** receives the service package compiled by the SDK. They deploy and manage the service in its operational state by addressing the **Infrastructure Providers** at the bottom, to lease the necessary compute, storage and network capacity. The economic viability of a network service improves greatly if these virtualized resources can be optimally scaled to fit real-time performance needs, without any noticeable interruption for the service users. The operational cost would also further decrease if the service is controlled and managed automatically. NFV and SDN have proven added value regarding resource virtualization and automated network control, and the SDK should assist in integrating these technologies. Moreover, vendors have proprietary knowledge about how the service internally works, while operators have their own private systems to deploy and manage the service. The SDK offers a way to bridge this gap by facilitating the interfacing between the involved parties and allowing closer cooperation during the service's lifetime.

## 2.1 The MANO Platform

Service Operators need to have an adapted deployment system to support the dynamicity enabled by NFV/SDN. Such a system is the **Management and Orchestration** (**MANO) platform**, as can be seen on Fig. 3. In accordance with the ETSI[1] defined framework for management and orchestration [5], the high-level functional blocks are these:

- The **NFV Orchestrator (NFVO)** maintains a global overview of the service topology. It calculates the placement and orchestrates the (scaled) VNFs and network links onto the available infrastructure. In the described use-case, the NFVO would decide where the Cache and DPI VNFs would be placed in the available access networks.
- The **VNF Manager (VNFM)** controls the lifecycle events of a single VNF such as instantiation, configuration and scaling. The VNFM of the Router in the selected use-case, could decide to scale in or out according to the required traffic rate.
- **Virtualized Infrastructure Manager (VIM) adapters** provide the NFVO and VNFM an interface to control the compute and storage nodes. Specialized VIMs control the network between the different infrastructure nodes or cloud data centers: a centralized SDN-controller for example, to setup the required links between VNFs in different access networks.

The communication between the different modules in the MANO platform happens by using pre-defined messages over a message broker, or the modules address each other's API directly. Additional features related to monitoring and automated healing of network services are also possible parts of the MANO framework [6].

Some of the automated NFVO and VNFM control functionalities, such as VNF startup and shutdown, can be quite generic. Other VNF lifecycle events, such as configuration, updating, migration or scaling, are likely to include very service-specific logic, custom-built by the developer. Therefore MANO platforms would need to plug-in customized control functions, shown in Fig. 3. Each service has its own NFVO and each VNF has its own VNFM. Suppose a scale-out would be required as traffic increases: The scaled-out topology is calculated by the specific VNFM, orchestrated by the service's NFVO and deployed using the involved VIM adapters. A more detailed explanation of the MANO framework is out of the scope of this article, but it is important to understand that the modular approach and split-up of the various service management and control features must be taken into account during service development.

---

[1] ETSI, NFV research group, https://irtf.org/nfvrg, accessed on 27 June 2017

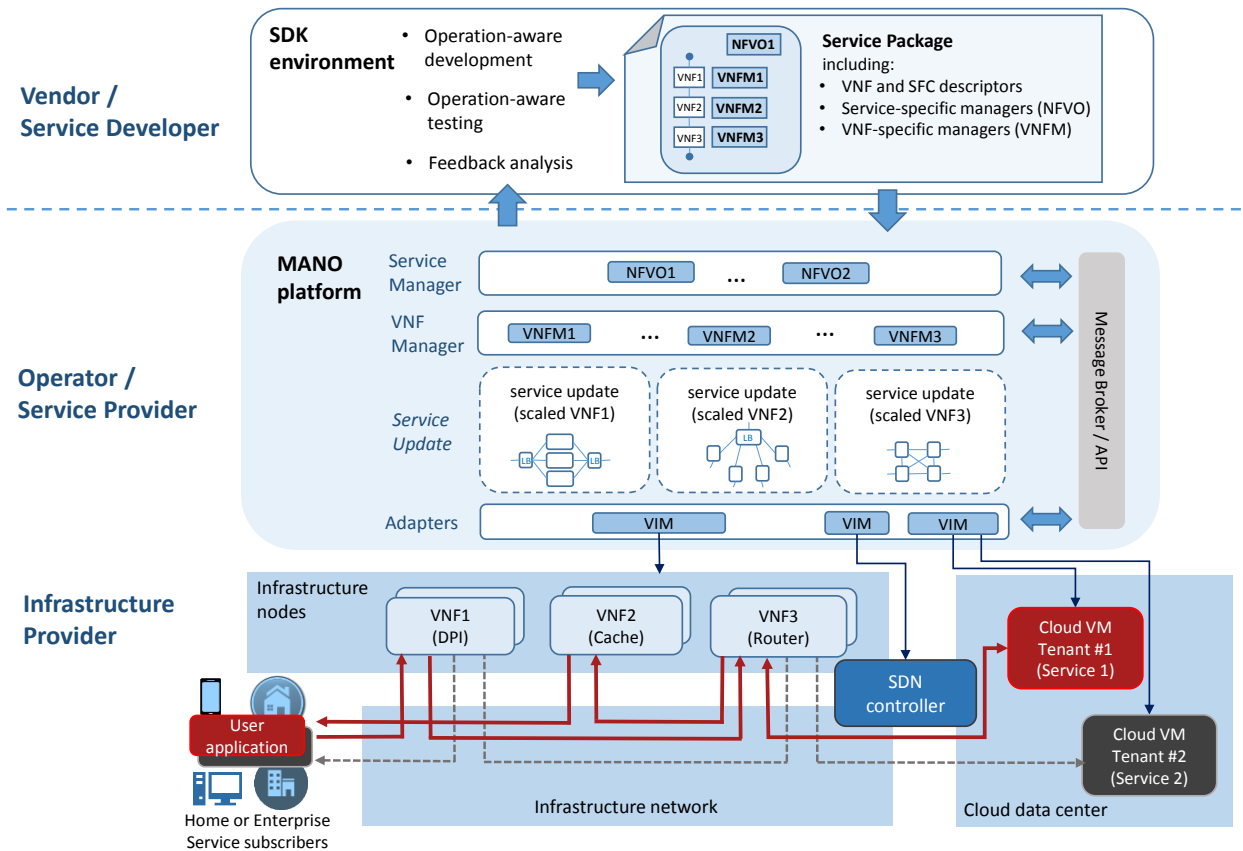

Figure 3: The SDK produces a Service Package which describes the chained VNFs in the network service and the required functions for customized management and control. This is deployed through the Service Provider's MANO platform.

## 2.2 The Service Package

To abstract the wide range of deployment and operational aspects of an SDN/NFV-based service, a programming model is helpful [7]. Essentially, network services can be seen as graphs, like the SFC example in Fig. 2. The VNFs are the nodes which can be enriched with annotations such as their resource requirements (number of CPU cores, amount of memory and storage), or other requirements like high availability. The edges are the links in the infrastructure network, specified by necessary bandwidth or the maximum delay, further constraining the placement in the physical infrastructure. Other abstractions, like network resiliency, can be mapped to redundant link configurations for example. Several flavors of such a model are being devised by ETSI-NFV [5] and several research projects such as UNIFY, T-NOVA, OSM and SONATA. Also open-source initiatives like TOSCA and OpenStack/HEAT have own models.

The service package includes everything needed to deploy the service in the operator's environment and bridges the boundary between vendors and operators. Figure 3 shows that the package should at least consist of:

- References to the actual VNF images to deploy on the infrastructure.
- A service graph that describes how the VNFs in the service are connected.

- All service or VNF specific logic in the form of NFVOs or VNFMs that can be plugged into the operator's platform.

Optionally, a definition of the expected feedback from the operator can be added. This can include a set of metrics to be monitored or certain alarms, triggered by a given threshold. Also shown in Fig. 3 is how multiple, in parallel deployed, service packages support a multi-tenancy scenario. In our example a second service using only DPI and Router VNFs is added, reaching a second VM in the cloud data center. The modules of the service package are developed by the vendor, and the operator should deploy all components on the infrastructure while respecting the constraints defined in the service package. It can be seen that service abstraction into packages allows vendors and operators to work in much closer collaboration, with still enough room for proprietary knowledge on either side. The role of the SDK is to support the creation and validation of this service package.

### 2.3  Telco-grade DevOps

The softwarized nature of SDN/NFV-based services, makes them a good fit for DevOps processes. A well-known methodology from the IT-world for building and maintaining software projects, but now applied to a collaboration between network service developers and telecom operators [4, 6]. At a high level, it resembles the "design for manufacturing" engineering concept, where the design facilitates the manufacturing process [1]. But in a telco-grade solution, the design should facilitate the operator's deployment. As explained above, the service package allows the operator to deploy the modular service on its own MANO platform. By using the SDK, the vendor or service developer has the ability to do operation-aware development and testing: After deployment, monitoring data can be analyzed to detect failures and debug any VNF or service-related functionality. As shown in Fig. 4, the service can re-iterate through the SDK, where it is edited and packaged again with any needed updates. This also enables continuous integration and continuous deployment (CI/CD), another common practice in software development. CI/CD merges development with testing, allowing to build code collaboratively and automatically check for issues. Figure 4 also depicts that an execution environment can be chosen from the SDK, so the service can be checked in parallel to production. The SDK features are detailed in the next section.

## 3  Introducing SDK Features for SDN/NFV - based Services

In the previous section we have explained how the service package enables an open interface between vendors and operators. As depicted in Fig. 4, we use the service package to implement two main categories of SDK features:  **formal pre-deployment checks** and a **functional verification** of the service. The toolset allows a developer or vendor to fully validate service updates and minimize the risk of failures, before deployed in production.

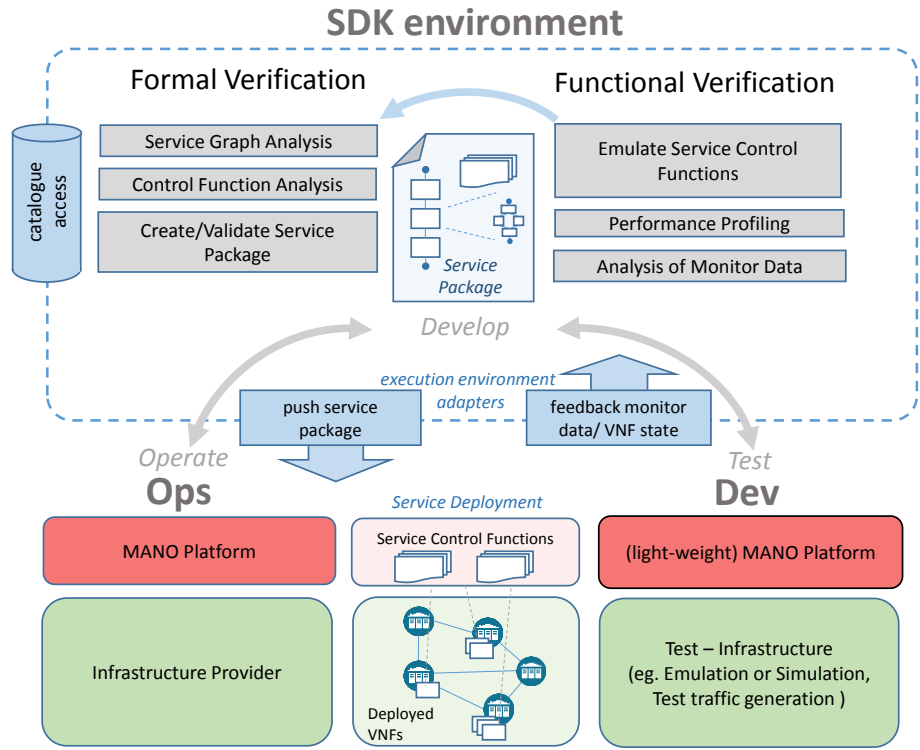

Figure 4: The SDK can edit and verify different parts of the service package before initiating service deployment in various execution environments.

## 3.1 Formal Verification Methods

By using formal verification methods, the service package can be logically checked for correctness, to make sure that the MANO platform is able to accept and deploy the network service. **Service Graph Analysis** can report issues such as invalid connection points, repeated paths and existence of cycles in the forwarding graphs which provide hints about the service integrity and deployability. The severity of a found issue may also depend on the VNF's capabilities, e.g. a cycle in a forwarding graph may not be problematic if an involved VNF is able to steer traffic and avoid endless loops. Possible bottlenecks for network congestion can be detected by analyzing the requested bandwidth for each link in the service and the performance specifications of the VNFs. If the SDK is tightly coupled to the targeted MANO platform, a formal **Control Function Analysis** can verify the code correctness or malicious code presence in any custom VNFM and NFVO. For example, it might be required that they are based on provided templates or parent classes, to make sure that they can be plugged into the MANO platform used by the operator. Practically, formal methods can be integrated into editors for generating network service descriptions. They can also be reused at the MANO entry point, to validate incoming service packages.

Previously validated functions or templates can be made available via a **Catalogue** for easy access by the SDK and integration into new services. Different data modelling languages can be used to define a service description (e.g. XML, JSON, YANG, YAML). With an associated schema, formal verification of the descriptor itself can be done. The integrity is further examined by ensuring all VNF references and images are accessible. Finally, the service model is packaged by the SDK, meaning that all required information to deploy the service is compiled into one single entity. By **pushing the service package** to the MANO platform, it is not needed to replicate all deployment mechanisms inside the SDK. For added security the package can be digitally signed, to trust that it was created by a known user, and to check if the package file was altered. The same key pair used for signing the package can also be used for user management in the execution environment to authenticate and authorize the SDK user. This adds a level of security to verify if the pushed package can be trusted or not.

## 3.2 Functional Verification Methods

As shown in Fig 4, the SDK has one main southbound interface: an adapter to **push the service package** to multiple execution environments. On one hand, deployment goes through the operator's MANO platform (e.g. leasing operation-qualified hardware configurations or operating system versions). On the other hand, a local emulation or simulation environment [8-10] can be used as a sandbox to test the service. This is likely to be less performant, but can be a cheaper and easier alternative for quickly checking basic functionality, to try configuration settings or generate test traffic in (parts of) the network service.

To gain operational insights, **monitor data can be received** through a second interface of the SDK platform. This interface can also be used to query the internal state of a VNF. Monitoring agents or traffic-generating test VNFs could be inserted at any location in the service, by simply updating the service graph. Mathematical techniques from regression analysis or machine learning can be used to process the monitored metrics through **Performance Profiling** of the VNF [11]. The performance profile enables predictable VNF performance and optimized resource usage. Automated scaling functions manipulate the service graph, as exemplified in Fig. 3 by load-balancing, hub-and-spoke or full-mesh topologies which can be seen as templates where any VNF can be plugged-in to. The SDK could provide these templates to the developer for integration into a custom scaling algorithm. If the scaled topology is combined with the VNFs' profiling data, the performance of the scaled topology can be estimated before it is deployed.

With a minimized pay-per-use cost model, over-provisioning should be limited and scaling algorithms must be well tested and fine-tuned. The **Emulation of Service Control Functions** in the SDK can reveal for example bugs in the service-control functions (VNFM, NFVO) that cause an exponential cost increase because too many resources are requested. To investigate placement algorithms, the SDK is not required to implement the full orchestration. It would be sufficient to check the NFVO function's graph output, showing onto which infrastructure nodes the VNFs in the service are mapped. A graph visualization in the SDK, could for example be helpful to evaluate the outcome of these algorithms. As shown in Fig. 4, the graph output of the service's control functions can in fact be fed back into the SDK's formal analysis tools for verification.

## 3.3 SDK Use-Case: Horizontal and Vertical Scaling

We revisit the SFC example of Fig. 2 to investigate how an SDK environment could support this service. Different aspects are highlighted in Fig. 5: **Custom horizontal scaling** is implemented on the Router VNF, which aggregates most of the service traffic and will therefore be more prone to changing loads (such as temporarily popular streams or night/day differences). The router VNF is made elastic, it changes the number of dedicated data-plane servers to forward the traffic. After deploying the updated topology, the forwarding tables in the new data-plane servers are configured accordingly. Details on the implementation can be found in [12]. A **local emulation environment** [10] is used to audit the scaling algorithm under test-traffic. By monitoring the throughput rate and packet loss of the elastic router, the scaling procedure can be validated. Additionally, updated placement or **custom orchestration** is implemented by the NFVO. More or less DPI and Cache VNFs can be part of the service, in function of the location and amount of service users. By checking the service graph modifications, the custom NFVO algorithm can be verified in the SDK.

The **vertical scaling capabilities** of the Cache VNF are also examined in Fig. 5. Using the automated profiling features demonstrated in [13], the throughput of the Cache VNF is tested under increasing CPU resource allocation. After statistically analyzing the time-series data, the SDK generates a more comprehensible way to describe the performance of the VNF. This helps to define the resource boundaries for a specific VNF performance. The performance thresholds can be filled in a VNFM function that will assign the required resources and optionally re-configure the VNF to use the newly allocated resources. Also, while testing the VNF under different resource allocations, implementation issues can be discovered [8] e.g. Is multi-threading support working correctly? Is the performance scaling linearly while adding resources?

This use-case exemplifies how the SDK should assist in validating the custom VNFM or NFVO control functions that can be plugged into the MANO platform. Using an execution environment, configuration and operation is functionally tested. At the end of the development process, a fully validated service package with highly automated control functions can be handed over to the operator/MANO platform.

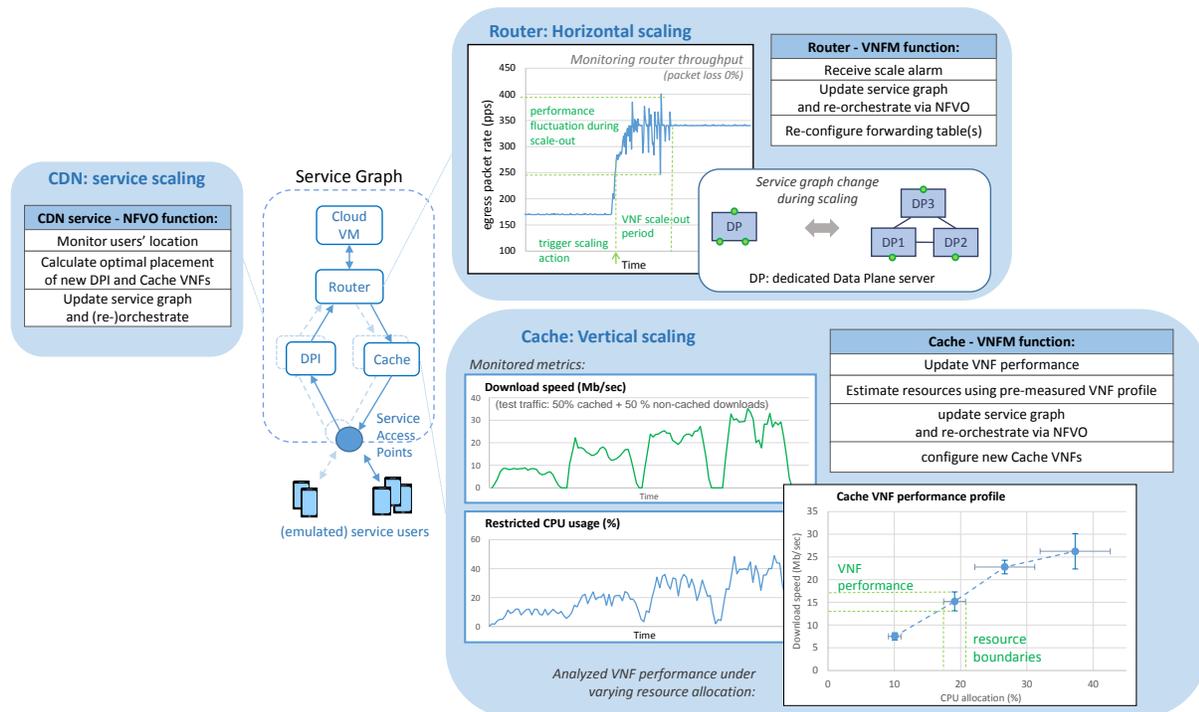

Figure 5: Custom control functions for the CDN service that can be created and validated using the SDK.

### 3.4  Limitations in Existing SDK Environments

Future NFV-based service deployments must tackle a wide spectrum of use-cases [2] (e.g. the 5G, IoT or Telco related service space). Customizable management and control is therefore an inevitable aspect of the service platform and should be supported from the design-phase onwards. Existing SDKs focus primarily on cloud applications, which have a very different nature compared to network services. Major commercial cloud providers follow a simplified model in their SDK. (i) One part of the toolset focuses on the application development: operating system or programming language specific. (ii) Another part helps the developer to connect to the data center, deploy the application and monitor it. Some cloud platforms support the execution of custom event handling code through the use of certain hooks inside the application[2]. There is however no test-environment available to trigger and validate the custom event handlers, without fully deploying the service.

Distinct open-source tools implement a specific part of the envisioned SDK environment, such as monitor data analysis [12, 13]. Adapters for light-weight or specialized environments [8-10] can deploy chained virtual machines or containers for testing or emulate the execution of service control functions on large scale data center topologies. A federated testbed uniting different technologies and administrators, is described in [14]. This shows the practical implementation of a distributed test-infrastructure across multiple owners.

Existing NFV technology overviews [2] can be updated with new concepts of customizable service management, implemented by recent open-source projects. The focus shifts to telco-grade NFV, beyond cloud provider functionality as mentioned earlier. Both UNIFY [3] and SONATA [4] define a MANO architecture that allows custom, service-specific control functionality, like scaling, configuration and placement. UNIFY has a SP-DevOps toolkit [6] for post-deployment troubleshooting and SONATA provides a pre-deployment tool-chain to describe, validate and package complete service chains [13]. OSM[3] and ONAP[4] are MANO platforms devised by industry-driven consortia, and have a clear roadmap towards an NFV-related SDK. Like SONATA, they both offer a set of design-time tools for easy service graph editing and packaging. ONAP additionally implements a Policy Subsystem that allows the creation of easily-updatable conditional rules, executed by ONAP's own control, orchestration, and management functions. A formal validation can detect policy conflicts. This framework, as explained in [15], implies however that the execution of the policy rules is embedded into ONAP and not modifiable by the service developer. The VNF management is enhanced by OSM and Open-Baton[5] with a set of libraries to implement a VNFM with specific VNF configuration scripts and an interface to the NFVO. Recent initiatives such as OPNFV and NGPaaS focus on facilitating NFV development and deployment across multiple open source eco-systems. Neither of the described platforms has already full development support for integrating customized service control.

## 4  Challenges and Opportunities

To avoid the need for bulky and unified development software, we proposed a limited but specialized feature-set for the SDK, built around the service package, as illustrated in Fig.4. This development environment is extended with key functional verification tools, based on feedback analysis of monitor data and VNFM/NFVO output. The main features are summarized in Table 1. We see however some challenges to fit the SDK into the real-world telco eco-system:

- A DevOps mindset between operators and vendors should be cultivated, bringing the Ops environment closer to development.

---

[2] Implementing custom lifecycle events can be done by e.g. Cloudify Lifecycle Events, AWS Lifecycle Hooks, Azure Functions webhooks, and service control using Google Cloud App Engine.
[3] Open-Source-MANO (OSM): https://osm.etsi.org/ (Accessed on 11 Sept 2017)
[4] The Open Network Automation Platform (ONAP): https://www.onap.org/ (Accessed on 11 Sept 2017)
[5] Open-Baton: Open-Source MANO framework: http://openbaton.github.io/ (Accessed on 11 Sept 2017)

- The same service package formats should be supported by the operator's MANO platform and the SDK. This would require a consolidation of several existing service package formats or the need for multiple package format translators.
- Security risks should be mitigated by authorization and authentication of the SDK user when pushing to the operator's environment. Possible exploits in the service control functions must be detected before deployment e.g. infinite scaling, unauthorized access to resources.
- Workflows for generic VNF tasks (orchestration, networking, start, update, terminate) should be defined and guaranteed by the MANO platform, taken out of the hands of the SDK.
- VNFs should be developed with elasticity in mind.

The proposed SDK environment creates however many opportunities to optimize the service lifecycle:

- Modeling and packaging the service leads to **easier validation**, **re-usability** and (re-)**deployment speedup**.
- Monitoring and profiling tools allow a **reliable reproducibility** and definition of the service performance.
- Creating customized placement and scaling algorithms enable a more **optimized resource usage**.
- Implementing highly automated management functions **decreases operational cost**.
- Supporting a clear service package format **lowers the barrier between vendors and operators.**

| Key Components | SDN/NFV related aspect | SDK features |
|---|---|---|
| **Service Package** | **Network Functions (VNFs)** | -Support for NFV specific program languages, APIs or libraries<br>-VNF state verification |
| | **Networking control** | -Support for SDN specific program languages, APIs or libraries<br>-Visualization of the network state |
| | **Service programming model** | -Service/VNF catalogue<br>-Re-usability of service templates<br>-Formal pre-deployment check<br>-Compilation into an easy-to-deploy service package<br>-Graphical verification of customized service graphs |
| **Custom service control functions (NFVO/VNFM)** | **VNF configuration** | -Programming support for NFVO/VNFM functions, tightly coupled to the MANO platform.<br>-Sandbox/emulator environment to test the VNF interfaces |
| | **Custom scaling** | -Simulation of scaling triggers<br>-Customization of VNFM templates for high-availability (auto--scaling, load-balancing)<br>-Verification of custom state migration procedures |
| | **Custom placement** | -Simulation of VNF orchestration<br>-Verification of NFVO output<br>-Visualization of the deployed service graph mapped on the available infrastructure |
| **Supporting functions** | **Monitored data** | -Packet stream analysis<br>-Data analytics (Regression Analysis, Machine Learning)<br>-Generation of custom test traffic and monitoring VNF/service metrics |
| | **Performance profile** | -Generation of a reliable VNF performance profile<br>-Capacity estimation and optimized resource planning<br>-VNF Benchmarking |

Table 1: Specific aspects of SDN/NFV-based services, whose development is assisted by new SDK features.

As advances in SDN and NFV help to transform telecom services into software-based network function chains, it is important that development tools keep up with this evolution. The holistic setup of the SDN/NFV-based service implies that different artefacts can be part of the service package, including elastic scaling mechanisms and possible resource and placement constraints. It requires a generic service programming model that is not yet standardized. While building further on DevOps principles and existing NFV architectures, we identified new SDK features to streamline the development and deployment of modern virtualized telecom services. We hope the presented development flow can give further directions to the ongoing research in SDN/NFV-based service creation.

# 5  Acknowledgement

This work has been performed in the framework of the SONATA and NGPaaS projects, funded by the European Commission through the Horizon 2020 and 5G-PPP programs. The authors would like to acknowledge the contributions of their colleagues of the projects consortia.

Supporting the End-to-End Lifecycle of NFV-based Telecom Services." In IEEE NFV-SDN2017, the IEEE Conference on Network Function Virtualization and Software Defined Networks, 2017. (Accepted, to be published in the conference proceedings on 8 Nov 2017.)

[14] Berman, Mark, Jeffrey S. Chase, Lawrence Landweber, Akihiro Nakao, Max Ott, Dipankar Raychaudhuri, Robert Ricci, and Ivan Seskar. "GENI: A federated testbed for innovative network experiments." *Computer Networks* 61 (2014): 5-23.

[15] ONAP-ECOMP AT&T Technology and Operations, "Ecomp (enhanced control orchestration management policy) architecture white paper," 2016. Online Available: https://about.att.com/content/dam/snrdocs/ecomp.pdf (Accessed on 11 Sept 2017)

# 7 Author biographies

**Steven Van Rossem** received a M. Sc. in Electrical Engineering in 2010 from K.U. Leuven (Belgium). After a five-year period working in the telecom industry, he started a PhD with the IDLab, imec research group at Ghent University in 2015. His research targets software-defined networking and network function virtualization, focusing on elasticity and performance profiling of network functions/network services. This work contributed to European research projects such as UNIFY, SONATA and NGPaaS.

**Wouter Tavernier** received a M.S. in computer science in 2002, and a Ph.D. degree in computer science engineering in 2012, both from Ghent University. He joined the IDLab, imec research group of Ghent University in 2006 to research future Internet topics. His research focus is on software-defined networking, network function virtualization, and service orchestration in the context of European research projects such as TIGER, ECODE, EULER, UNIFY, and SONATA.

**Didier Colle** is a full professor at Ghent University. He received a Ph.D. degree in 2002 and a M.Sc. degree in electrotechnical engineering in 1997 from the same university. He is group leader in the imec Software and Applications business unit. He is co-responsible for the research cluster on network modelling, design and evaluation (NetMoDeL). This research cluster deals with fixed Internet architectures and optical networks, Green-ICT, design of network algorithms, and techno-economic studies.

**Mario Pickavet** is professor at Ghent University since 2000 where he is teaching courses on discrete mathematics, broadband networks and network modelling. He is leading the research cluster on Network Design, Modelling and Evaluation, together with Prof. Didier Colle. In this context, he is involved in a large number of European and national research projects, as well as in the Technical Programme Committee of a dozen of international conferences.

**Piet Demeester** is a professor at Ghent University and director of IDLab, imec research group at UGent. IDLab's research activities include distributed intelligence in IoT, machine-learning and datamining, semantic intelligence, cloud and big data infrastructures, fixed and wireless networking, electromagnetics and high-frequency circuit design. Piet Demeester is a Fellow of the IEEE and holder of an advanced ERC grant.